# Parallel bandit architecture based on laser chaos for reinforcement learning

**Takashi Urushibara[1], Nicolas Chauvet[1], Satoshi Kochi[2], Satoshi Sunada[2], Kazutaka Kanno[3], Atsushi Uchida[3], Ryoichi Horisaki[1] and Makoto Naruse[1]**

[1] *Department of Information Physics and Computing, Graduate School of Information Science and Technology, The University of Tokyo, 7-3-1 Hongo, Bunkyo-ku, Tokyo 113-8656, Japan*

[2] *Graduate School of Natural Science and Technology, Kanazawa University, Kakuma-machi, Kanazawa, Ishikawa 920-1192, Japan*

[3] *Department of Information and Computer Sciences, Saitama University, 255 Shimo-Okubo, Sakura-ku, Saitama City, Saitama 338-8570, Japan*

## Abstract

**Accelerating artificial intelligence by photonics is an active field of study aiming to exploit the unique properties of photons. Reinforcement learning is an important branch of machine learning, and photonic decision-making principles have been demonstrated with respect to the multi-armed bandit problems. However, reinforcement learning could involve a massive number of states, unlike previously demonstrated bandit problems where the number of states is only one. Q-learning is a well-known approach in reinforcement learning that can deal with many states. The architecture of Q-learning, however, does not fit well photonic implementations due to its separation of update rule and the action selection. In this study, we**

organize a new architecture for multi-state reinforcement learning as a parallel array of bandit problems in order to benefit from photonic decision-makers, which we call parallel bandit architecture for reinforcement learning or PBRL in short. Taking a cart-pole balancing problem as an instance, we demonstrate that PBRL adapts to the environment in fewer time steps than Q-learning. Furthermore, PBRL yields faster adaptation when operated with a chaotic laser time series than the case with uniformly distributed pseudorandom numbers where the autocorrelation inherent in the laser chaos provides a positive effect. We also find that the variety of states that the system undergoes during the learning phase exhibits completely different properties between PBRL and Q-learning. The insights obtained through the present study are also beneficial for existing computing platforms, not just photonic realizations, in accelerating performances by the PBRL algorithms and correlated random sequences.

## 1. Introduction

Recently, photonic approaches to accelerate artificial intelligence have been intensively studied [1], ranging from optical fibre-based neuromorphic computing [2], on-chip deep learning [3], photonic reservoir computing for time series prediction [4] to exploit the unique physical attributes of photons for supervised learning. Reinforcement learning is another important machine learning technique to maximize long-time rewards in unknown environments [5]. Photonic principles have been proposed to efficiently resolve the difficult trade-off between the search actions (exploration) for better choice and the exploitation of the accumulated knowledge in the multi-armed bandit (MAB) problem [6]. There are a variety of physical resources to be used for such decision-making problems ranging from wave-particle duality of single photons [7], chaotic oscillation of lasers [8], unstable oscillation in ring lasers [9], entangled photons [10], among others.

However, the abovementioned studies have been limited to bandit problems in which the number of states is only one [6]. Conversely, the number of states in the case of reinforcement learning in general can be huge [11–14], indicating that further insights are indispensable to examine the potential of photonics to real artificial intelligence issues. Meanwhile, current photonic decision-makers for bandit problems take account of the unique attributes of photons as well as physical and technological constraints. Therefore, a novel composition is essential to upgrade the capability of the photonic reinforcement learning system for multi-state situations while benefiting from the unique properties of photons. This paper proposes a photonic reinforcement learning architecture consisting of a massive array of MAB solvers, which we call Parallel Bandit Architecture for Reinforcement Learning, or PBRL in short. We demonstrate PBRL with chaotically oscillating laser time series.

In conventional Q-learning [11–13], a so-called Q-table is required to determine the most appropriate action for a given state; the number of states gives the number of rows of the Q-table, whereas the number of possible actions determines the number of columns of the Q-table. Each element of a Q-table is called Q-value, which stands for the estimation of the total future reward.

Learning proceeds by updating Q-values. In each state, the agent chooses the action which gives the maximum Q-value. However, this simple action selection may not provide sufficient exploration when the agent is still learning the appropriate Q-table; therefore, one needs to employ an action selection policy separately from Q-learning. The ε-greedy method is a representative example, where the agent sometimes chooses actions randomly rather than based on the Q-table to enforce exploration. Such a basic structure is schematically illustrated in figure 1(a). We consider that such conventional Q-learning architecture does not fit well with photonic implementation because of the separation of the update rule and the action selection. On the other hand, photonic decision-making developed for the bandit problems so far is structurally very simple [7,8]; for example, it can solve two-armed bandit problems by a single parameter that corresponds to a threshold level of photon detection in the case of chaos-based decision-maker [8].

PBRL also requires a table, but it is not a Q-table; here, we regard the learning process of PBRL as updating an array of threshold values, each of which is compared with chaotic time series. Hereafter, we call the array of thresholds a threshold table. The action is chosen after comparing the chaotic time series and the threshold. In other words, we regard multi-state reinforcement learning as an array of bandit problems where the individual bandit problem decides actions in a different single state. After an action, the thresholds, or the threshold table, are updated depending on the reward and the state transition. Herein, the threshold table is in charge of *both* action exploration and exploitation as schematically represented in figure 1(b), which is clearly different from Q-learning (figure 1(a)) where exploration and exploitation are separated.

A variety of studies are being conducted in reinforcement learning from quantum information processing standpoints [15–20]. In particular, Steinbrecher *et al*. examined quantum optical neural networks for reinforcement learning, wherein the cart-pole balancing problem was solved [15]. Saggio *et al.* discussed quantum speed-up in multi-agent reinforcement learning [16]. Flamini *et al.* examined photonic architecture for reinforcement learning using single photons to solve GridWorld problems [17]. While these studies focus on utilizing quantum attributes of photons for reinforcement learning, PBRL in this paper does *not* depend on the quantum attributes of photons. Ultimately, the interest of PBRL is to exploit the ultrafast operation capability of laser chaos [21]; random number generation [22,23], reservoir computing [4], and bandit solvers [8] are utilizing such properties. Furthermore, it is noteworthy that the latest silicon CMOS electronics allow ultrafast operations over 100 GHz range with respect to analog signal processing [24,25], which is strongly demanded by recent beyond 5G and 6G telecommunications technologies. The simple architecture of PBRL could meet such ultrafast circuit technologies. We elaborate on the associated topics in section 4.

In the meantime, the parallel bandit architecture by the present study can be numerically implemented in the machine learning platforms in use today. Moreover, acceleration is accomplished by correlated time series such as laser chaos. That is to say, the insights triggered by photonics approaches to reinforcement learning can be transferred back to the existing computing systems with enhanced performance, which is another contribution of this study.

In this paper, we first generalize the photonic architecture for bandit problems toward reinforcement learning problems involving multiple states as a parallel array architecture. As an example, we deal with a cart-pole problem in the discussion. We demonstrate that the negative autocorrelation inherent in chaotic laser time series improves the adaptation performance of PBRL, which is validated by the comparison with surrogate chaos time series; this is one unique attribute derived by chaotic lasers unlike pseudorandom numbers. We show that PBRL completes the learning process faster than Q-learning. Moreover, we find that learning progresses in a completely contrasting manner between PBRL and Q-learning by analyzing the variety of states that the system undergoes during the early stages of the learning process.

## 2. Architecture

### 2.1. Q-learning

We start by reviewing conventional Q-learning where an agent acts in a stochastic environment. Here we assume that an agent moves over a finite number of states given by $N$. The number of possible actions that can be performed by the agent is called $M$. An action selection induces a state transition. Since we assume that the environments are Markovian, the state transition depends only on the current state and the chosen action. In most cases, reinforcement learning aims to maximize the total reward of the agent after certain consecutive actions by some estimation of the total reward, which sometimes employs the notion of time discount $\gamma$ regarding the reward obtained at each step.

Although Q-learning updates the Q-table as learning progresses, we remark that the reward setting, or parameters associated with rewards, must be decided prior to the learning. In other words, it is often the case that the reward setting exists as a hyperparameter to be tuned, which is a non-trivial issue. Furthermore, one needs to specify how the agent chooses each action; this is because Q-learning itself only determine the rule to update Q-values. In the present study, we adopted the ε-greedy method [12,13] as a well-known and conventional strategy. The details of Q-learning and the ε-greedy method are described in the appendix.

## 2.2. Parallel bandit architecture for reinforcement learning

As mentioned briefly in the introduction section, chaotic laser time series has been successfully utilized for solving two-armed bandit problems, by which faster adaptation to unknown environments has been demonstrated [8]. Therein, the binary action (Action 1 or Action 2) is selected by comparing the incoming signal level of chaotic laser intensity level and the threshold; if the chaos signal is larger than the threshold, take Action 1, otherwise Action 2. The learning proceeds through threshold adjustment; for example, if Action 1 does yield rewards, let the threshold decrease so that the likelihood of selecting Action 1 increases in the subsequent chance of actions. Similarly, if Action 1 does not provide rewards, let the threshold increase so that Action 1 is less likely to be chosen in the next chance of actions. Since the incoming signal is a chaotically oscillating time series, for example, Action 2 can be selected even when the threshold is low.

PBRL in the present study extends such a threshold adjustment approach to multi-state reinforcement learning. For simplicity, we assume the total number of actions in each state is two (Action 1 or Action 2). Each state is associated with a two-armed bandit problem; $TH_j$ denotes a threshold variable for the state $s_j$. In the state $s_j$, the agent compares the threshold value $TH_j$ with the value of chaotic laser time series at the index of this state. The action is immediately determined; if the chaos value is larger than $TH_j$, take Action 1; otherwise, Action 2. After an action, the agent moves to the next state according to the dynamics of the environment and receives rewards. Based on the state transition and the rewards, threshold variables are updated; the revisions may regard not only the threshold of the current state but also the states that the agent has experienced until the current state. Hence, Q-table in Q-learning becomes an array of threshold values or the threshold table. The rule to update the threshold table needs to be configured according to the learning task. The formula for the cart-pole balancing problem is described later in the results and discussion section.

What is significantly different between the former genuine two-armed bandit problem and the present PBRL is that a two-armed bandit for a certain state could be related to other multiple states. Therefore, multiple thresholds variables can be updated simultaneously depending on the reward and the state transition. Also, it should be emphasized that while Q-learning needs an extra strategy for random selections for exploration, PBRL is inherently equipped with such exploration ability thanks to the chaotic bandit structures. That is, exploration and exploitation are unified in PBRL.

Finally, in order to avoid potential confusion or misunderstanding, here we would like to emphasize that the proposed PBRL concerns a single-agent system, not a multi-agent one. That is, the present study is *not* about parallel reinforcement learning or multi-agent learning systems such as reference [14] in the literature. In this study, a single agent utilizes the parallel array of bandit solvers.

## 2.3. Demonstration

We simulated a one-dimensional cart-pole balancing problem, which is schematically shown in figure 1(c), to examine the principle of PBRL, its performance, and the impact of chaotic time series. We used the OpenAI Gym framework [28] for the numerical study. There are four variables to describe the states of the system; the lateral displacement of the cart, the velocity of the cart, the angle of the pole, and the angular velocity of the pole. We extracted a valid range of these four variables and divided each of them into six divisions; hence there are in total $N = 6^4 = 1296$ different states. The action has two alternatives; to push the cart to the right or the left by a fixed mechanical force ($M = 2$). The failure of an action is defined as the case when the cart's displacement or the pole's angle is out of the specified region. If an action does not lead to immediate failure, we call this action successful. In the initial state, the cart is located almost exactly at the centre with almost zero velocity, and the pole stands almost completely vertically and thoroughly stationary. The error from the centre and the stationary differs in every trial. The number of actions by the agent is represented in the unit of *step*, while consecutive actions and state transitions from the initial state until failure are defined as an *episode*.

Here we describe the threshold adjustment mechanism for this cart-pole balancing situation in PBRL. Assume that the state of the system is $s_j(t)$ at step $t$. As described above, the agent decides the action based on the comparison between the signal level of chaos and the threshold $TH_{j(t)}$. Let us denote the action conducted at time $t$ by $a(t)$. If the action is successful, the agent shifts the threshold $TH_{j(t)}$ by a fixed value of $\Delta TH$ to the direction in which the agent becomes more likely to choose the same action in the subsequent chance of action at this state.

On the other hand, if the action fails, the agent looks up the past states $s_j(t')$ ($1 \leq t' \leq t$), which the agent has experienced since the initial state of the episode. The agent also looks up their associated actions $a(t')$ conducted at states $s_j(t')$ at steps $t'$ ($1 \leq t' \leq t$). The agent updates the thresholds for these states $s_j(t')$ by $A_0 \gamma^{t-t'}$ as a penalty, where $A_0$ is a constant positive number and $\gamma$ accounts for time discount, which is given by a real number in the range of $0 < \gamma < 1$. Here is an example situation. The agent took the action $a(t')$ as a result of the level comparison that the chaos signal level is *larger* than the threshold $TH_{j(t')}$ at time $t'$. Then, later at time $t$, the action fails. In this case, the agent increases the threshold corresponding to the state taken at time $t'$ by the penalty given by $A_0 \gamma^{t-t'}$. The amount of penalty is heavier when $t'$ is closer to $t$ (failure point) with the intention that the recent actions may be strongly relevant to the failure. In this study, $\Delta TH$, $A_0$, and $\gamma$ were configured as summarized in table 1. The details are described in the Supplementary Information.

To summarize the above principle, the update rule of PBRL is as follows. We assume that the agent is at state $s_{j(t)}$ at step $t$ and takes action $a(t)$. If the action taken is Action 1, $a(t)$ is equal to 1, and if it is Action 2, $a(t)$ is equal to 2. If the agent succeeds, only threshold $TH_{j(t)}$ is updated by

$$TH_{j(t)} \leftarrow TH_{j(t)} + (-1)^{a(t)} \, \Delta TH. \tag{1}$$

If the agent fails, all the thresholds which the agent has experienced over the past are updated by

$$TH_{j(t')} \leftarrow TH_{j(t')} - (-1)^{a(t')} \, A_0 \, \gamma^{\,t-t'} \qquad (1 \leq t' \leq t). \tag{2}$$

As described above, a single episode is defined as a set of steps from an initial state until it fails. In the simulation, we defined 150 steps as the maximum number of consecutive successes in an episode. Hence, when the agent does not fail through all 150 actions since the initial state, this particular episode completes without any penalty. The agent repeats an episode from a slightly different initial state without resetting the $TH$ values through 1,000 episodes. We call such 1,000 consecutive episodes a single learning round.

As described shortly below, we compared the learning performances of PBRL coupled with chaotic laser time series, normally distributed pseudorandom numbers, and uniformly distributed random numbers. The chaotic laser time series were experimentally obtained by a conventional time-delayed feedback architecture using a semiconductor laser [8]. As shown in figure 2(a), the chaotic laser time series possess a negative autocorrelation, exhibiting its minimum value at the time lag of 50 ps. In reference [8], it was demonstrated that chaotic laser time series resolves the two-armed bandit problem faster than other random sequences such as uniformly distributed pseudorandom numbers. Furthermore, the fastest adaptation was realized by the chaotic laser time series sampled with a 50 ps interval, which exhibits the maximum negative autocorrelation.

In the numerical evaluation, we conducted 9,000 learning rounds for each of 10 different sampling intervals of the chaos; they consisted of 300 different chaotic laser time series combined with 30 sets of 1,000 different initial states. Further detailed conditions are described in table S1 of the Supplementary Information. The blue, green, and yellow curves in figure 2(b) indicate the histogram of the signal level distributions normalized by the summation of the incidences; therefore, the vertical axis can be interpreted as the probability density.

## 3. Results and discussion

### 3.1. Learning speed

We evaluated the number of continuous successful steps in each episode to examine the learning speed. Figure 3 summarizes the time evolution of the average steps of continuous successes over learning rounds by PBRL equipped with laser chaos time series and various pseudorandom numbers as well as that by Q-learning. In other words, we quantify how the number of successful steps, or equivalently, at what steps the agent fails, evolves as the episodes progress. See the Supplementary Information regarding the details of the parameter settings. One minor remark here is that the plots in figures 3(a) and 3(b) display only the first 62 episodes, while the actual learning process lasted 1000 episodes.

Figure 3(a) shows the *continuously successful steps* over learning rounds by PBRL utilizing various random numbers and by the conventional Q-learning. Here we utilize (i) the original laser chaos time series, (ii) a random shuffle surrogate of the chaos time series, (iii) normally distributed random numbers, and (iv) uniformly distributed random numbers for PBRL. Here, the surrogate (ii) is a randomly time-shuffled version of the original chaotic laser time series, making the correlation being zero. The blue, red, green, and yellow curves in figure 3(a) show the number of continuous successes as a function of episodes when the time series is given by (i), (ii), (iii), and (iv), respectively. We observe that the uniformly distributed random numbers are slower than the other three methods. Meanwhile, the other three methods do not exhibit evident differences. This slowness of uniformly distributed random numbers matches the result in reference [8]; the uniformly distributed random numbers are the slowest in the experiment of the bandit problem using the threshold adjustment method.

For our PBRL approach, the impact of the inherent temporal dynamics of chaos becomes an interesting issue. Figure 3(b) examines the adaptation speed of PBRL with different sampling intervals of the original chaotic laser time series. More specifically, the sampling interval is configured differently from 10 ps to 100 ps in a 10-ps interval, and the resultant learning performances are analyzed. For simplicity, figure 3(b) shows five of the continuously successful steps evolutions when the sampling interval is configured by 10, 30, 50, 70, and 90 ps. We observe that the blue curve, corresponding to the sampling interval of 10 ps, is inferior to the others. For further comparison, we take the averages of the continuously successful steps from the 11[th] to 20[th] episodes, during which learning progresses well, and the exploration-exploitation dilemma is serious. The blue curve in figure 3(c) exhibits its peak values when the sampling interval is about 50 ps. Such a property coincides with the autocorrelation shown in figure 2(a), which shows the maximum negative at 50 ps, indicating the impact of temporal dynamics in laser chaos on the resultant reinforcement learning. Indeed, with the surrogate chaos time series, the learning performance becomes almost flat as denoted by the red curve in figure 3(c), validating the impact of the time-domain nature of the chaos herein. Meanwhile, the green and yellow dashed lines in figure 3(c) denote the performances when normally and uniformly distributed pseudorandom numbers are utilized, respectively. Since these pseudorandom numbers do not meet the concept of sampling intervals, constant values are marked throughout the horizontal range.

Finally, we compared PBRL with Q-learning combined with an ε-greedy algorithm. See the appendix for the details of Q-learning and the ε-greedy method. The four parameters for ε-greedy Q-learning have been optimized as summarized in table 1(b); the details are described in the Supplementary Information. The orange curve in figure 3(a) shows the result by Q-learning, whereas the other four curves represent PBRL combined with chaotic time series, surrogate chaos time series, normally distributed pseudorandom numbers, and uniformly distributed pseudorandom numbers. We clearly observe that the conventional method is slower than PBRL.

## 3.2. Variety of states

PBRL realizes a larger value of continuously successful steps with a fewer number of episodes than Q-learning, as observed in figure 3(a), meaning that a faster adaptation is observed by PBRL. To examine the underlying mechanisms, we shed light on aspects other than the evolution of the learning performance; here, we focus on the agent's experiences by introducing the concept of the *variety of states*. The definition of the variety of states is as follows. First, we record the states which the learning process undergoes during every ten episodes. We remark that although there are in total $N = 1296$ states, the learning actually may go through much fewer states than 1296. We evaluate the number of experienced states every ten episodes and take the average of all of the learning trial rounds, which is the definition of the variety of states. A larger number of the variety of states is considered to be relevant to a broader exploration.

Figures 4(a) and 4(b) demonstrate the time evolutions of the variety of states when chaotic laser time series and its random shuffle surrogate are utilized for PBRL, respectively. The five curves therein correspond to different sampling intervals: 10, 30, 50, 70, and 90 ps. The analysis itself was conducted with sampling intervals ranging from 10 ps to 100 ps with an interval of 10 ps. From figure 4(b), we observe that all curves follow nearly the same trace, whereas the curves in figure 4(a) exhibits diversity depending on the sampling interval. These are another manifestation of the effects originating from the autocorrelation inherent in chaotic lasers.

Furthermore, we take the average of the first two plots for the detailed analysis; the first plot corresponds to the epoch from the first episode to the tenth episode, while the second does to the epoch from the eleventh to the twentieth. In these epochs, learning of PBRL is considered to be highly active, as we saw in figure 3(a). The blue curve in figure 4(c) shows the results as a function of the sampling interval, where it exhibits its minimum peak when the sampling interval is 50 ps, which corresponds to the maximum negative in the autocorrelation of the laser chaos time series. Conversely, the red curve in figure 4(c) concerns the surrogate laser chaos time series, which shows negligible dependencies on the sampling intervals. That is, here we observe the correspondence between the variety of states and the autocorrelation of laser chaos. If we interpret a large number of the variety of states during the early episodes as the wider exploration, we can claim that the negative autocorrelation may shorten the duration of the exploration. Indeed, if we invert the upper and lower sides of figure 4(c), it becomes almost the same as figure 3(c). From such discussions, we can draw an insight that different random numbers provide different exploration manners, leading to different adaptation speeds of learning.

Figure 4(d) compares the variety of states of PBRL with Q-learning. The orange curve, corresponding to Q-learning, exhibits a completely different trace compared with the proposed method, which is marked by the other four curves. PBRL explores extensive states in the early episodes and then shrinks them to limited states. On the contrary, Q-learning begins the learning phase with a few states at first and then gradually expands the transition range. Meanwhile, we should remark that although we may change the exploration degree of Q-learning by tuning the parameter $\varepsilon_0$ in the $\varepsilon$-greedy algorithm, we have to decide it prior to the task execution. On the other hand, PBRL adjusts the amount of exploration automatically by the threshold adjustments; thresholds around the centre mean weighting on exploration, while biased thresholds indicate focusing on exploitation. Therefore, we claim that this built-in mechanism to balance the ratio of exploration and exploitation is one crucial factor in the quick learning made possible by PBRL.

### 3.3. Remarks for future

Finally, we discuss a few remarks for future studies. First, the interpretation of the variety of states may be studied further. With a deeper understanding of the variety of states, we may clarify the fundamental features of PBRL.

Second, we will examine further applications other than the cart-pole balancing problem, which accompanies only two choices of actions. Scaling the number of actions is an important issue. Indeed, scalable decision-making principles have been demonstrated using chaotic time series up to 64-arms based on time-domain multiplexing of chaotic laser time series [29]. The applicability of such principles to our photonic method aiming at multi-state situations is an interesting future topic. Moreover, the present cart-pole situation is a simple and deterministic problem. Roughly speaking, once the machine continuously successes 150 actions, or one whole episode, it only needs to reproduce its actions. In this sense, the current settings could easily reach a ceiling for reward maximization and have less complexity for exploration. A stochastic environment may be a more challenging problem to exploit the potential abilities of the chaos-based method.

Third, we can consider extending the present architecture to multi-agent systems, even with maintaining the simple cart-pole balancing as the problem under study. Indeed, Chauvet *et al.* examined the use of entangled photons to benefit from quantum correlations to maximize total rewards [10]. Furthermore, Amakasu *et al.* succeeded in extending the number of arms in collective decision making by utilizing the orbital angular momentum of photons [30]. Combining the present approach with quantum ones [10,15–20,30] to induce collective effects is also an interesting future study.

The fourth point is discussions on ultrafast operations of PBRL. As remarked in section 1, the present architecture is motivated by utilizing ultrafast properties of laser chaos [21–23]. Furthermore, the latest advancements in silicon analog circuits [24,25] would bring the fusion of electronics and photonics, as already observed in terahertz photonics [31]. Indeed, the design and analysis of comparator or thresholding circuits to deal with chaotic time series [32] and history memorization circuits [33] have been theoretically and numerically investigated. It should be remarked that the cart-pole balancing problem studied in this paper is a simple benchmark problem in reinforcement learning. Examining truly beneficial scenarios and applications utilizing ultrafast aspects is one of the future studies.

Finally, this paper shows that the algorithm based on PBRL provides faster adaptation than Q-learning as well as additional acceleration by correlated time series. This indicates that the PBRL approach contributes to the existing hardware platforms based on CPU. Indeed, recently, Okada *et al.*

demonstrated a theory to explain the positive impact of correlated time series on bandit problems based on the correlated random walk model [34]. Actually, this study [34] was originally triggered by the question of why negatively correlated laser chaos time series accelerates the decision making [8,35]. Consequently, such obtained insights through the mathematical model lead to benefits that we can experience in existing computing architectures. Photonics-based considerations for computing eventually provide novel perspectives.

## 4. Summary

We propose a reinforcement learning principle involving many states in which every state is regarded as a bandit problem solved by threshold adjustment coupled with chaotic laser time series; we call the proposed methods parallel bandit architecture for reinforcement learning or PBRL in short. Using a cart-pole balancing problem as an instance, we demonstrate that PBRL realizes faster adaptation than Q-learning. Furthermore, PBRL accomplishes faster adaptation when triggered by chaotic time series than uniformly distributed pseudorandom numbers. We identify the negative autocorrelation as one of the main advantages of PBRL using laser chaos time series by observing the disappearance of the original properties when we utilize a randomly shuffled laser chaos time series.

Furthermore, we introduce the concept of the variety of states, which quantifies how the learning process undergoes versatile experiences. With this indicator, we first find the apparent effects of the autocorrelation inherent in chaotic laser time series, implying that the negative autocorrelation may shorten the exploration duration of PBRL. Second, we observe a clearly contrasting nature between PBRL and Q-learning; while the variety of states in PBRL begins with a large number at early episodes and gradually decreases toward the stable states, Q-learning starts with a small number of the variety of states, gradually increases and converges to a similar value of PBRL. This difference manifests the ability of PBRL to explore in a short duration. This work paves a new way for the application of chaotic time series, photonic architecture and technologies for reinforcement learning as well as sheds new light on the utilities of physical processes on intelligent functionalities.

## Acknowledgments

The authors thank Andre Roehm for the discussion of the study. This work was supported partly by the CREST project (JPMJCR17N2) funded by the Japan Science and Technology Agency and Grants-in-Aid for Scientific Research (JP20H00233) funded by the Japan Society for the Promotion of Science.

## Appendix

### Laser chaos

The chaotic laser time series we used in this research was the same as the reference [8].

### Q-learning

The agent updates what is called the Q-value denoted by $Q(s, a)$, which indicates the expected total reward after the agent takes action $a$ at the state $s$. The revision of $Q(s_t, a_t)$ is based on the renewal formula

$$Q(s_t, a_t) \leftarrow Q(s_t, a_t) + \alpha \left[ r_{t+1} + \gamma \left\{ \max_a Q(s_{t+1}, a) \right\} - Q(s_t, a_t) \right], \tag{3}$$

where $\gamma$ and $\alpha$ are parameters for time discount and learning rate, respectively [11]. The two-dimensional array of Q-values whose columns and rows correspond respectively to actions and states is called Q-table. Q-learning means the way to update the Q-table; the decision of actions needs to be specified separately, such as the ε-greedy method [5], softmax decision-making model [36], or roulette wheel selection [37].

**Reward setting for Q-learning**

As represented in equation (3), Q-learning requires a predetermined reward setting. In the present study, the reward setting for Q-learning was specified as follows. After an action, if it does not fail immediately, the agent receives a reward of +1; if it fails before the 145[th] step in the episode, the agent gets $-r_{\text{penalty}}$ as a reward. Here $r_{\text{penalty}}$ is a positive parameter that must be tuned. Details of the tuning are described in the Supplementary Information.

**Greedy method**

In the greedy method, the agent chooses the action that maximizes $Q(s, a)$ at time $t$, which means the agent focuses on exploitation rather than exploration. If the agent wants to attach importance to exploration more, he or she can adopt the ε-greedy method, which is described below.

**ε-greedy method**

In the ε-greedy method, the agent chooses an action randomly with a probability of $\varepsilon_t$, while with the probability of $1 - \varepsilon_t$ the agent chooses the action that maximizes $Q(s, a)$ at time $t$. The probability $\varepsilon_t$ monotonically decreases toward zero as the time elapses. In this study, time $t$ corresponds to the number of episodes that the agent goes through, and we employed the formula $\varepsilon_t = \varepsilon_0 / (t + 1)$ where $\varepsilon_0$ is a fixed parameter. If $\varepsilon_0$ is equal to 0, the ε-greedy method is equal to the greedy method.

**The setting of parameter tuning**

PBRL using chaotic laser time series contains three parameters in adjusting the array of threshold variables: (1) The adjustment amplitude of the threshold when the action is successful, denoted by $\Delta TH$, (2) The adjustment amplitude when the action fails, which is denoted by $A_0$, and (3) the time discount denoted by $\gamma$. Q-learning contains four parameters: (1) The penalty of action failure when a consecutive 145-step of successful actions is not realized, denoted by $r_{penalty}$, (2) The time discount $\gamma$, (3) The learning rate $\alpha$, and (4) the exploration rate $\varepsilon_0$, which is associated with the ε-greedy algorithm. In optimizing these parameters, we used the Nelder-Mead method [38] for both methods. For PBRL, the Golden-section search [39] is additionally utilized.

As discussed in the main text, we examine the impact of four kinds of different random sequences combined with PBRL. As a reminder, the four kinds of random sequences are the following: (i) laser chaos time series, (ii) its surrogate data, (iii) normally distributed pseudorandom numbers, and (iv) uniformly distributed pseudorandom numbers. It is noteworthy that the optimal parameters are different depending on the probability distribution of the random sequences subjected to the learning process. Thus, in the parameter tuning of PBRL, we first tune parameters using normally distributed pseudorandom numbers by the Nelder-Mead method. After this, we conduct fine-tuning for randomly shuffled surrogate time series, normally distributed pseudorandom numbers and uniformly distributed pseudorandom numbers by the Golden-section search. For the chaotic laser time series, we adopt the same parameters as the randomly shuffled surrogate time series because they have the same probability distribution.

Here, we describe the details of the fine-tuning for PBRL. After the tuning with normally distributed pseudorandom numbers by the Nelder-Mead method, we have a set of $\Delta TH$, $A_0$, and $\gamma$. The value of $\Delta TH$ or $A_0$ should vary according to the probability distribution because these two decides the scale of the thresholds which are compared with various scale of random numbers as described in figure 2(b) in the main article. On the other hand, we assume that the same $\gamma$ is optimal for the other random numbers, and the ratio of $\Delta TH$ and $A_0$ should be maintained among the other random sequences because these two are less likely to depend on probability distributions rather than the learning architecture which is common among different random numbers. Thus, after the Nelder-Mead method tuning step, we introduce the ratio $c$ and rewrite $\Delta TH$ and $A_0$ as $10^c \times \Delta TH$ and $10^c \times A_0$, respectively. In the fine-tuning, we tune the parameter $c$ by the Golden section search.

In order to run the Nelder-Mead method or the Golden section search, we need to decide the Figure of Merit (FOM). We think about the average evolution of the continuously successful steps in the Cart-Pole balancing simulations and define the FOM as the last episode which cannot succeed 145 steps continuously (See figure S1). Table S1 shows the number of learning rounds of the Cart-Pole simulations to calculate the average evolution. As shown in the table, this repeat number is different between the cases of the Nelder-Mead method and the Golden section search in order to reduce calculation time. One remark is that the number of episodes is limited by the maximum 1000; when the learning does not complete before 1000 episodes, the FOM is given by 1000.

As the definition of FOM implies, a small FOM indicates fast learning; therefore, the goal of the tuning is to minimize FOM. In addition, to save the calculation time, we predetermine FOM for *obviously wrong parameters* as 2000 without any actual calculation. In PBRL, the obviously wrong parameter ranges are defined as the cases

$$( \gamma \leq 0 \text{ or } \gamma \geq 1) \text{ or } ( \Delta TH \leq 0 \text{ or } \Delta TH \geq 10) \text{ or } ( A_0 \leq 10 \text{ or } A_0 \geq 1000)$$

while in Q-learning, they are defined as the case of

$$(r_{\text{penalty}} \leq 0 \text{ or } r_{\text{penalty}} \geq 1) \text{ or } (\gamma \leq 0 \text{ or } \gamma \geq 1) \text{ or } (\alpha \leq 0 \text{ or } \alpha \geq 1) \text{ or } (\epsilon_0 \leq 0 \text{ or } \epsilon_0 \geq 1).$$

**Parameter turning of PBRL**

As described above, we begin with the Nelder-Mead when normally distributed pseudorandom numbers are subjected to PBRL. The initial four parameter settings of $[\Delta TH, A_0, \gamma]$ are configured as follows:

$$[\delta_{\Delta TH}, 10 + \delta_{A_0}, 1 - \delta_\gamma], [\delta_{\Delta TH}, 1000 - \delta_{A_0}, 0.5], [10 - \delta_{\Delta TH}, 10 + \delta_{A_0}, 0.5], [\delta_{\Delta TH}, 10 + \delta_{A_0}, \delta_\gamma],$$

where $\delta_{\Delta TH} = 0.1$, $\delta_{A_0} = 1$, $\delta_\gamma = 0.01$.

After 20 iterations of a simplex transformation in the Nelder-Mead method, we obtain the minimum FOM and the corresponding three parameters given by

$$[\text{FOM}, \Delta TH, A_0, \gamma] = [45, 3.208, 547.0, 0.6774].$$

Then, the Golden-section search is adapted with $[\Delta TH, A_0, \gamma] = [10^c \times 3.208, 10^c \times 547.0, 0.6774]$. We tune the parameter $c$ respectively for the random shuffle surrogate time series, uniform random numbers, and normal random numbers for fairness. The initial four values of $c$ are commonly given, and they are given by $(2, 0, 2(\phi - 1), 2\phi)$ where $\phi = \frac{1+\sqrt{5}}{2}$. After 25 iterations, the values of $10^c$ for surrogate time series, normal random numbers, and uniform random numbers converge to 0.5508, 0.6705, and 1.522, respectively. These values correspond to the values of each parameter shown in  in the main article.

In the Golden section method, we utilize an original operation not to stop narrowing the range. The Golden section method holds four values of $c$ at one time; here, we call them $c_1$, $c_2$, $c_3$, and $c_4$ where $c_1 < c_2 < c_3 < c_4$. If $c_1$ has the same FOM of $c_4$ and the FOM of $c_2$ equals the FOM of $c_3$, we shrink 90% the range of $c$; that is

$$c_{\text{centre}} \leftarrow \frac{c_1 + c_4}{2},$$

$$c_1 \leftarrow c_{\text{centre}} + r\,(\,c_1 - c_{\text{centre}}\,),$$

$$c_2 \leftarrow c_{\text{centre}} + r\,(\,c_2 - c_{\text{centre}}\,),$$

$$c_3 \leftarrow c_{\text{centre}} + r\,(\,c_3 - c_{\text{centre}}\,),$$

$$c_4 \leftarrow c_{\text{centre}} + r\,(\,c_4 - c_{\text{centre}}\,)$$

where $r$ is given by 0.9.

**Parameter turning of Q-learning**

We use the Nelder-Mead method for the parameter optimization of Q-learning. We do not need the Golden section search because Q-learning does not have variations of random numbers. In the Nelder-Mead method, we configure the initial parameter sets of [ $r_{penalty}$, $\gamma$, $\alpha$, $\varepsilon_0$ ] as

$$[1000 - \delta_r, \delta, \delta, \delta], [1000 - \delta_r, 0.5, \delta, \delta], [\delta_r, 1 - \delta, 0.5, 1 - \delta], [\delta_r, \delta, 1 - \delta, 0.5 ], [\delta_r, \delta, \delta, \delta],$$

where $\delta_r = 1$ and $\delta = 0.01$. After 20 iterations of a simplex transformation in the Nelder-Mead method, we obtain the minimum FOM and the corresponding four parameter values of

$$[\text{FOM}, r_{penalty}, \gamma, \alpha, \varepsilon_0] = [148, 773.8, 0.8494, 0.2265, 0.4653],$$

which are also described in table 1(b) of the main article.

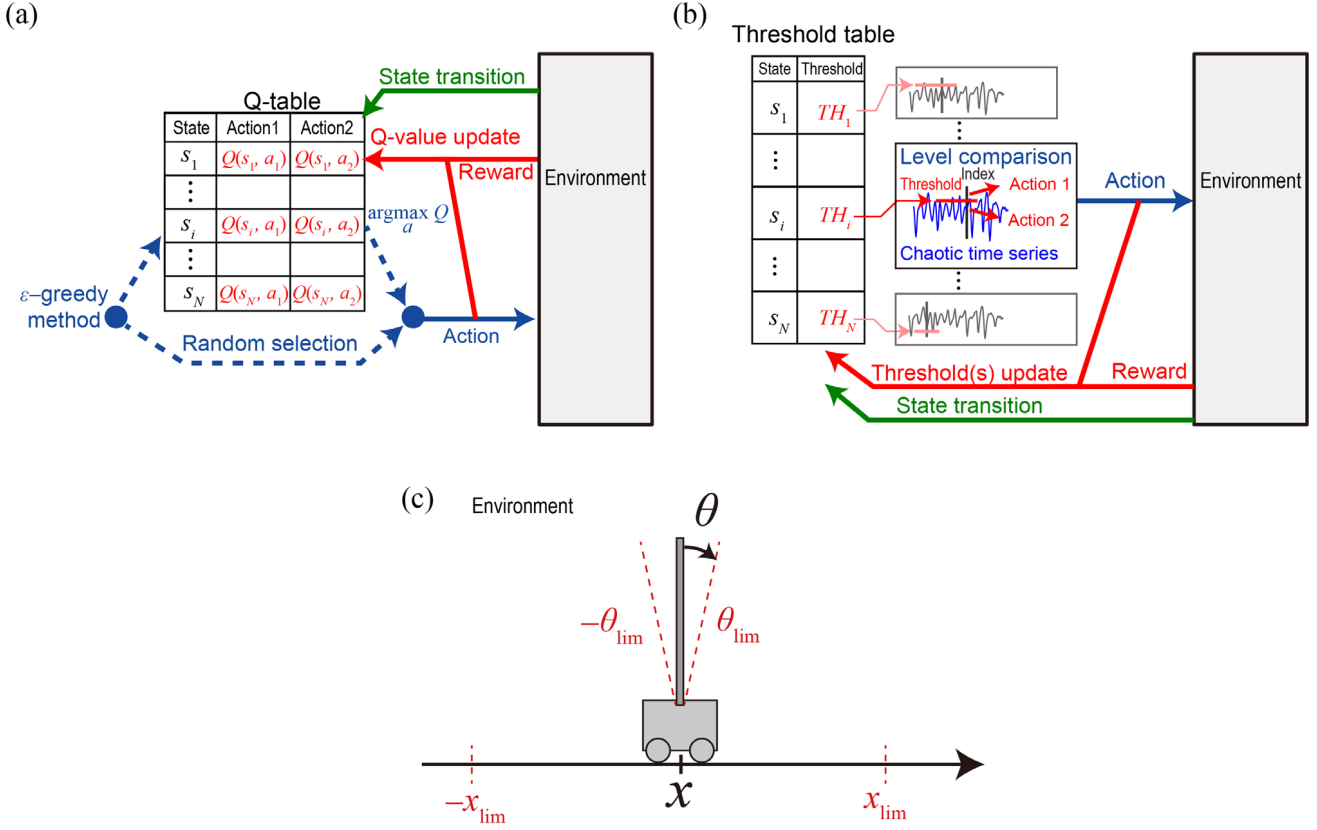

Figure 1. Parallel bandit architecture for reinforcement learning (PBRL). (a) The architecture of Q-learning combined with ε-greedy method and (b) PBRL. In (a), the object of learning is $Q(s, a)$, the expected value for total future reward after an action $a$ in a state $s$. In order to explore effectively, the agent sometimes chooses an action randomly; otherwise, the agent chooses argmax$_a Q(s, a)$ in the current Q-table (ε-greedy method). In (b), the object of learning is $TH_i$, which is compared with the value in the chaotic laser time series in order to decide the action. (c) The setting of the cart-pole balancing problem. We used OpenAI Gym for the simulation [28]. It is a one-dimensional balancing problem. The action has two alternatives: to push the cart to the right or to push it to the left by a force that has a fixed absolute value. The failure of an action is defined as the case when the cart's displacement or the pole's angle goes out of the predetermined region.

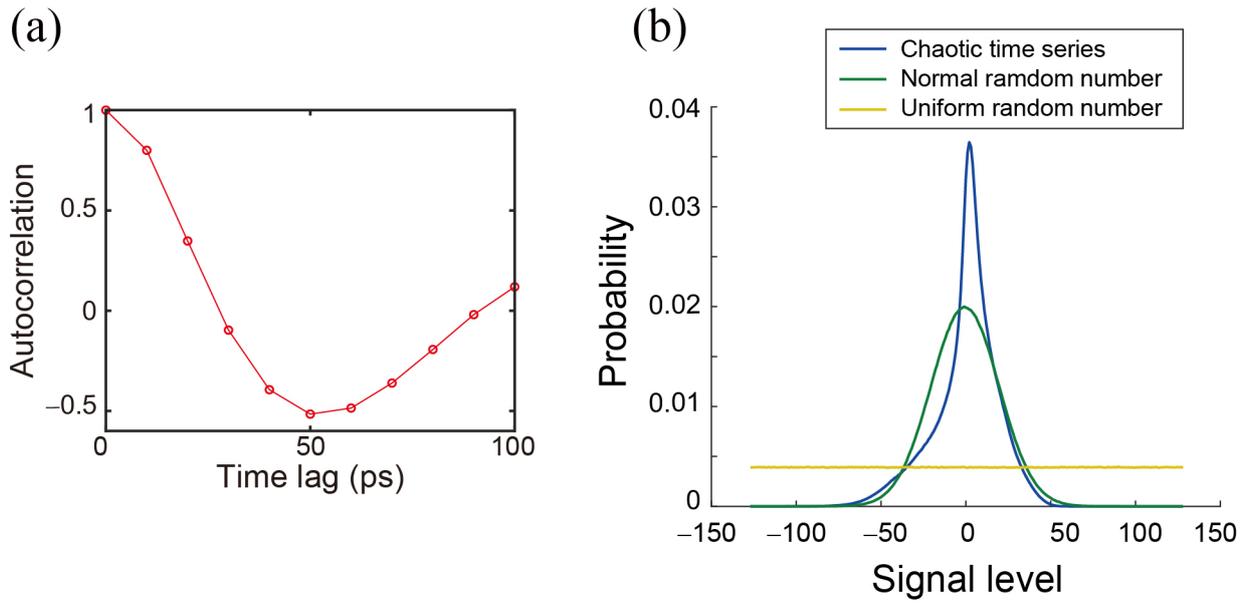

Figure 2. Properties of the chaotic laser time series. (a) Autocorrelation of chaotic laser time series. (b) Probability of the signal level of chaotic laser time series, normally distributed pseudorandom numbers, and uniformly distributed pseudorandom numbers. The signal level takes integer values between −127 and 128.

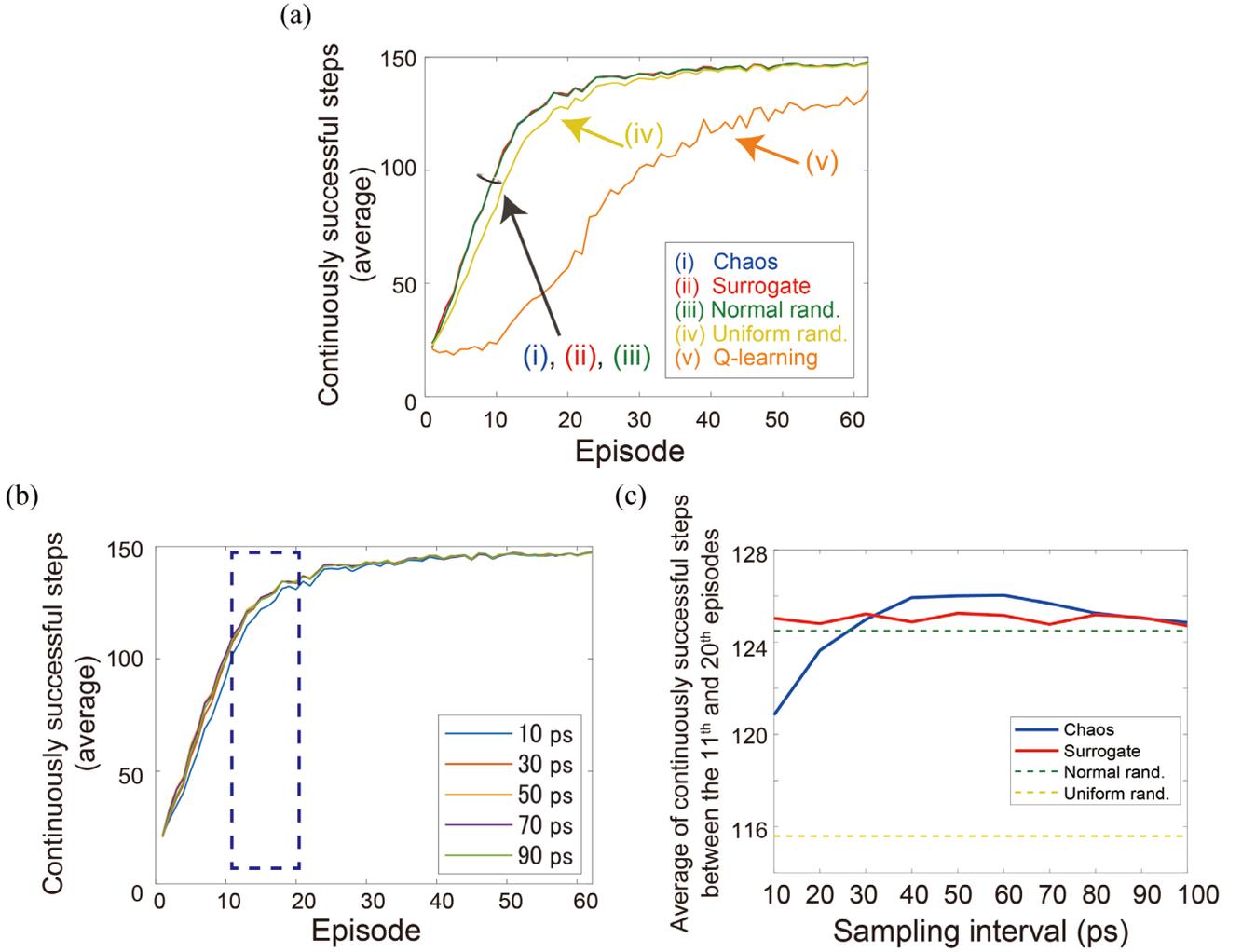

Figure 3. Learning performance evaluation. (a) Learning performance comparison of PBRL utilizing (i) chaotic laser time series, (ii) randomly shuffled surrogate laser chaos time series, (iii) normally distributed pseudorandom numbers, and (iv) uniformly distributed pseudorandom numbers, and (v) Q-learning. Here, (i) and (ii) display the average of 80,000 learning rounds, including ten variations of sampling intervals, while (iii), (iv), and (v) show the average of 8,000 learning rounds because they do not have variations of sampling intervals. (b) Learning performance comparison of PBRL with chaotic laser time series sampled by different intervals from 10 ps to 100 ps. (c) The average vertical axis value from the 11th episode to the 20th episode in (b) varied by different sampling intervals of chaos. Since normal random numbers and uniform random numbers do not have the concept of sampling intervals, we use the same value over every sampling interval.

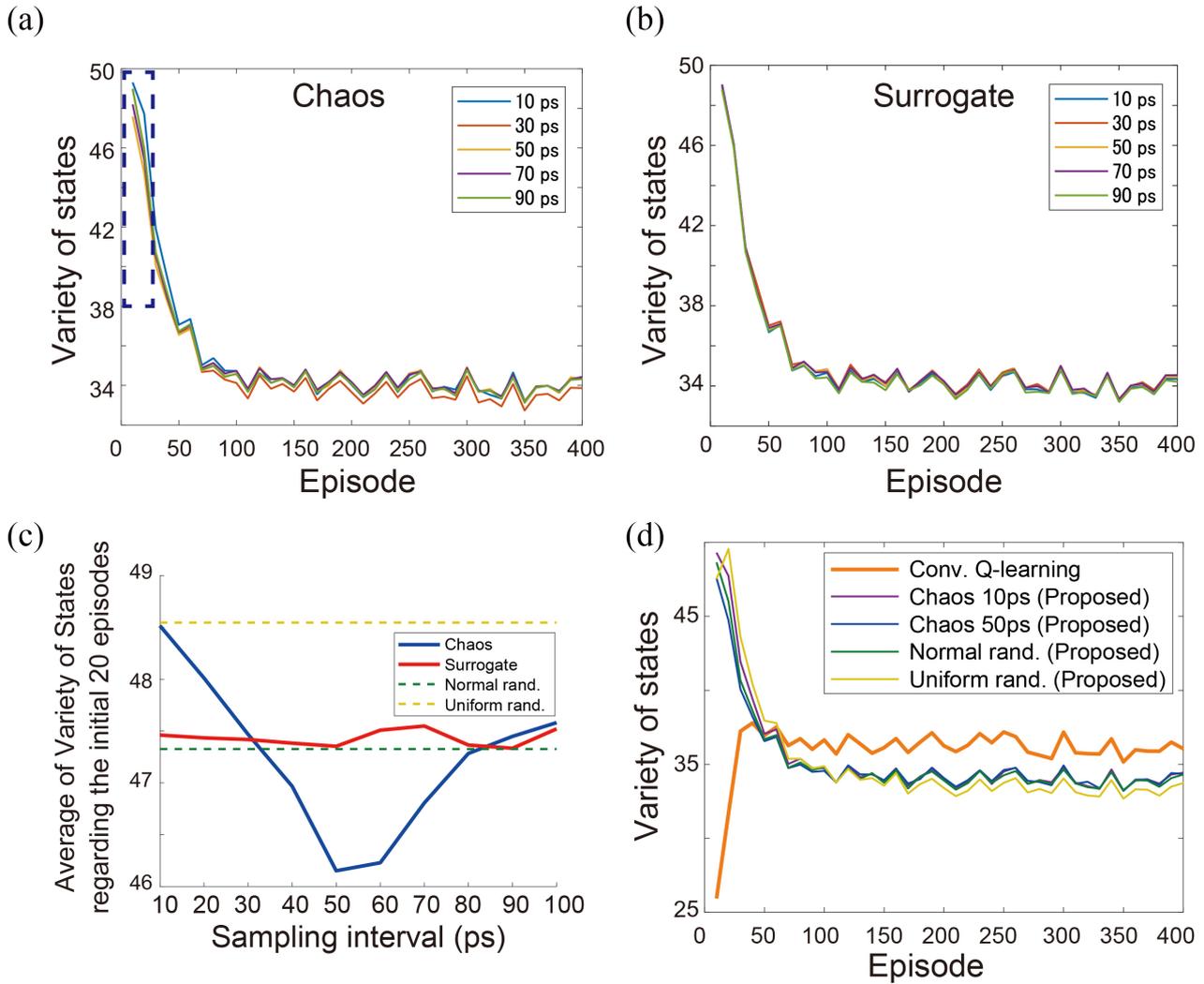

Figure 4. How the learning progresses in PBRL. (a,b) Time evolution of the variety of states, which is defined as the average number of experienced states in every ten episodes. (a) The variety of states for PBRL with chaotic laser time series sampled by different intervals to vary their autocorrelation. (b) The variety of states for PBRL with randomly shuffled surrogates of chaotic time series, whose autocorrelations are always zero. (c) The average value of the first two plots in the evolution of the variety of states varied by different sampling intervals of chaos. Since normal random numbers and uniform random numbers do not have the concept of sampling intervals, we use the same value over every sampling interval. (d) The variety of states for the results of PBRL compared by that of Q-learning.

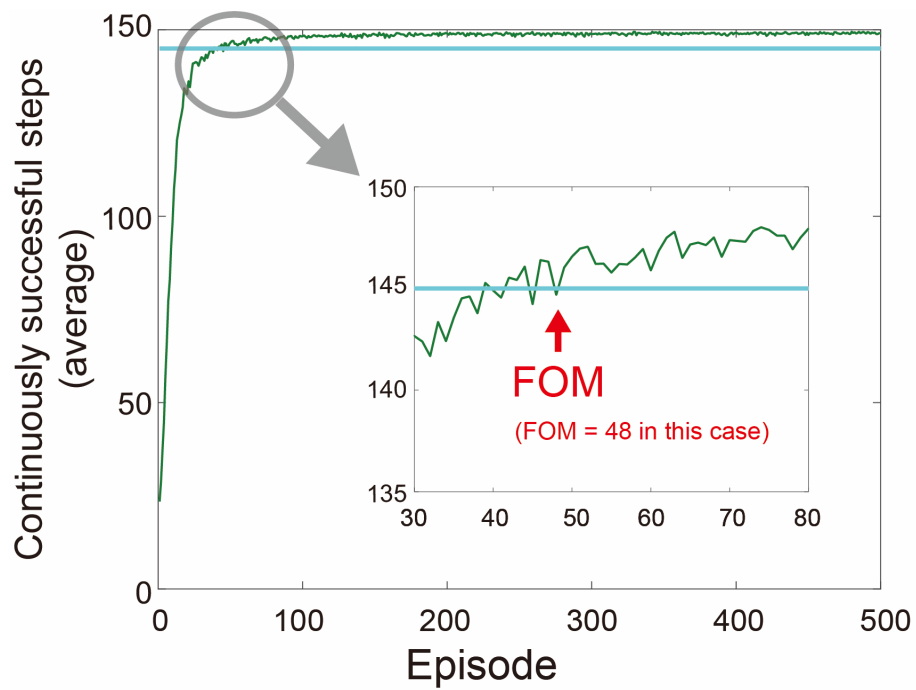

Figure S1. The definition of the FOM.

Table 1. Parameter settings in PBRL and Q-learning combined with ε-greedy method. The details of parameter tuning are described in the Supplementary Information.

(a) PBRL

| Time series | $\Delta TH$ | $A_0$ | $\gamma$ |
|---|---|---|---|
| Chaos / Surrogate | 1.767 | 301. 3 | 0.6774 |
| Normal random number | 2.151 | 366.8 | 0.6774 |
| Uniform random number | 4.881 | 832.5 | 0.6774 |

(b) Q-learning combined with ε-greedy method

| $r_{penalty}$ | $\gamma$ | $\alpha$ | $\varepsilon_0$ |
|---|---|---|---|
| 773.8 | 0.8494 | 0.2265 | 0.4653 |

Table S1. The number of learning rounds of the Cart-Pole balancing simulations to calculate FOM.

| | Initial states | Random number sequences | Total |
|---|---|---|---|
| Nelder-Mead | 40 | 120 | 4800 |
| Golden section | 30 | 90 | 2700 |